\documentstyle[12pt,epsf]{article}

\def\fun#1#2{\lower3.6pt\vbox{\baselineskip0pt\lineskip.9pt
        \ialign{$\mathsurround=0pt#1\hfill##\hfil$\crcr#2\crcr\sim\crcr}}}

\renewcommand\({\left(}
\renewcommand\){\right)}
\renewcommand\[{\left[}
\renewcommand\]{\right]}

\newcommand\ee{\end{equation}}
\newcommand\be{\begin{equation}}
\newcommand\eea{\end{eqnarray}}
\newcommand\bea{\begin{eqnarray}}

\newcommand\mpl{M_{\rm P}}

\newcommand\lsim{\mathrel{\rlap{\lower4pt\hbox{\hskip1pt$\sim$}}
    \raise1pt\hbox{$<$}}}
\newcommand\gsim{\mathrel{\rlap{\lower4pt\hbox{\hskip1pt$\sim$}}
    \raise1pt\hbox{$>$}}}

\def\dslash{\not{\hbox{\kern-2pt $\partial$}}}
\def\Dslash{\not{\hbox{\kern-4pt $D$}}}
\def\Oslash{\not{\hbox{\kern-4pt $O$}}}
\def\Qslash{\not{\hbox{\kern-4pt $Q$}}}
\def\pslash{\not{\hbox{\kern-2.3pt $p$}}}
\def\kslash{\not{\hbox{\kern-2.3pt $k$}}}
\def\qslash{\not{\hbox{\kern-2.3pt $q$}}}

 \newtoks\slashfraction
 \slashfraction={.13}
 \def\slash#1{\setbox0\hbox{$ #1 $}
 \setbox0\hbox to \the\slashfraction\wd0{\hss \box0}/\box0 }
 
\def\ee{\end{equation}}
\def\be{\begin{equation}}

\newcommand\sub[1]{_{\rm #1}}

\newcommand\Tr{{\rm Tr}\,}

\begin{document}

\begin{flushright}
LANCS-TH/9823
\\hep-ph/9812232\\
(Dicember 1998)
\end{flushright}
\begin{center}
{\Large \bf Hybrid inflation with running inflaton mass}

\vspace{.3in}
{\large\bf  Laura Covi}

\vspace{.4 cm}
{\em Department of Physics,\\
Lancaster University,\\
Lancaster LA1 4YB.~~~U.~K.}

\vspace{.4cm}
{\tt E-mail: l.covi@lancaster.ac.uk}
\end{center}

\vspace{.6cm}
\begin{abstract}
We realize and study a model of hybrid inflation in the context of 
softly broken supersymmetry. The inflaton is taken to be a flat direction in 
the superfield space and, due to unsuppressed couplings, its soft supersymmetry
breaking mass runs with scale. Both gauge and Yukawa couplings are 
taken into account and different inflationary scenarios are investigated
depending on the relative strenghts of the couplings and the mass spectrum.

\end{abstract}

\section{Introduction}

One of the well--known problems of implementing the usual slow--roll 
inflation picture\footnote{For a general discussion 
and references on inflation see \cite{ly98}.}
 in the framework of supersymmetric theories is the
fact that the second of the flatness conditions,
\bea
\epsilon &\equiv & {1\over 2} \( {V'\over V} \)^2 \ll 1\\
|\eta | &\equiv & \left| {V''\over V}\right| \ll 1,
\eea
where $V$ is the inflaton potential and we have set 
$\mpl \equiv (8\pi G)^{-1/2} \equiv 1 $, is usually violated 
by supergravity corrections \cite{cllsw}. 
All the scalar particles' squared masses (and therefore 
also the inflaton's) receive a contribution at least 
of order $V$ that spoils completely the flatness of the potential.

Few proposals exists in literature able to solve this problem without
fine tuning. We will in the following consider an inflationary model 
of the type proposed by Stewart \cite{ewanloop1,ewanloop2}, where
quantum corrections to the inflaton mass flatten a region of the potential.
The original model \cite{ewanloop1,ewanloop2}, studied also in 
\cite{clr98},
considered the case of a charged inflaton field, where the dominant 
corrections were due to the gauge coupling and neglected the role of 
the Yukawa coupling. 
Since the model is of the hybrid type, and the end of inflation is
determined by the Yukawa couplings, such a procedure has to be justified. 
We will see in an explicit case that
this approximation is indeed acceptable in the case of small Yukawa
coupling, but that in general such couplings can play a 
fundamental role in the flattening so that also the case of a
singlet inflaton is possible.

In the next section we will briefly review Stewart model with a
charged inflaton and introduce the formalism for taking into
account one loop quantum corrections; in section 3 we will describe 
an explicit model based on $SU(N)$ non-abelian gauge group.
In section 4 we will write the one-loop Renormalization Group
equations for the parameters involved. In section 5 we will
consider the case of very small Yukawa coupling, while in
section 6 and 7 the case of a Yukawa coupling respectively
of the order and much greater than the gauge coupling. In the 
limit of negligible gauge coupling, we will recover also the 
case where no gauge symmetry is present. 
Finally we will discuss the results and the naturalness
of the model and comment on the experimental signatures. 
We will not address in detail this last issue, but refer to 
\cite{clr98,cl98} for a model independent analysis.

\section{The running-mass model}

In the model proposed by Stewart \cite{ewanloop2}
(see also the review \cite{ly98}), slow-roll inflation occurs, 
with the following Renormalization Group improved potential
for the canonically normalized inflaton field $\phi$;
\be
V =V_0 + \frac{1}{2} m_\phi^2 (\phi) \phi^2 + 
 \frac{1}{2} m_\psi^2 (\phi) \psi^2   
 + \frac{1}{4} \lambda (\phi) \phi^2 \psi^2 
\cdots  \,.
\label{vinf}
\ee
The constant term $V_0$ is supposed to dominate at all relevant field 
values. Non--re\-nor\-ma\-li\-zable terms, represented by the dots, 
give the potential a minimum at large $\phi$,
but they are supposed to be negligible during inflation. 
The last two terms also vanish during inflation, since $\psi =0$, 
but are responsible for the exit from the inflationary period.
In fact at some critical value $\phi_c$, the $\psi$ field 
effective mass $  m_\psi^2 (\phi_c) + 1/2 \lambda (\phi_c) \phi_c^2 $
becomes negative and the fields roll towards the true vacuum 
with vanishing cosmological constant, characterized by 
\bea
\langle \phi \rangle &=& 0 \\
\langle \psi \rangle &=& V_0^{1/2}/|m_{\psi}|.
\eea

The inflaton mass-squared and all the other parameters
depend on the renormalization scale 
$Q$, and following \cite{ewanloop1,ewanloop2}
we have taken
\be
Q=\phi ,
\label{Qphi}
\ee
where now $\phi$ denotes the classical v.e.v of the inflaton field
during inflation.
Such choice for the renormalization scale minimizes the one loop
correction to the potential: in fact the main one loop contribution 
comes from the fields that acquire a mass proportional to 
$\phi $ and therefore goes like $\ln (\phi/Q)$
for $\phi$ larger than any other scale. If this is not the case 
some other $Q$ will be appropriate and the simplification given by 
eq.(\ref{Qphi}) is no more viable. We will assume that the
inflaton v.e.v. is the dominant scale up to the end of inflation.

At the Planck scale, $m_\phi^2 (M_{Pl})$ is supposed to be 
negative, with the generic magnitude
\be
|m^2_0 |=|m^2_\phi (M_{Pl}) | \sim V_0
\label{mexpect1}
\ee
coming from supergravity corrections \cite{cllsw,clr98}.

If there were no running, this would give $|\eta|\sim 1$,
preventing slow--roll inflation. But at field values below the
Planck scale, the RGE's drive $m^2_\phi (\phi)$ 
to small values, corresponding to $|\eta(\phi)|\ll 1$, and slow--roll
inflation can take place there\footnote{We are not addressing the
problem of initial conditions ad assume that the inflaton tunnels
in the small $\eta $ region from other values, for example the
minimum given by the non-renormalizable terms at large field values.}.
We have in fact
\bea
\epsilon &=& \phi \[m^2_\phi + {1\over 2} {dm^2_\phi\over d\ln\phi} \] 
\label{epsilon} \\
\eta &=& m^2_\phi + {3\over 2} {dm^2_\phi\over d\ln\phi}
+ {1\over 2} {d^2m^2_\phi\over d(\ln\phi)^2}
\label{eta}
\eea

Since in this model the $\eta$ parameter changes considerably
as $\phi$ decreases, slow-roll inflation is assumed to continue until 
some epoch $\phi\sub{end}$, when $\eta(\phi)$ becomes of order $1$.
Then $\phi$ start a brief phase of fast--roll up to the
critical value $\phi\sub c$, where the mass term of 
the other field $\psi$ becomes negative and inflation finally ends 
some number $N\sub{fast}$ of $e$-folds after the end of slow-roll
inflation, when the fields settle down in the true vacuum.

\section{The model}

Let us consider the case of the superpotential
\be
W = \lambda S \Tr \(\phi_1 \phi_2\)
\label{W}
\ee
where $S$ is a singlet chiral superfield, while $\phi_i$  are chiral superfield
in the adjoint representation of the gauge group $SU(N)$.
We can exclude other terms in the superpotential invoking some kind of 
discreet R-symmetry forbidding terms like $\Tr\(\phi_i^2\)$ or higher 
powers of $S$. One such example could be $W \rightarrow e^{i\alpha} W $,
$ S \rightarrow e^{i\alpha} S $, $ \phi_{1/2} \rightarrow e^{\pm i\alpha/4} 
\phi_{1/2}$.

In this case we can easily compute the scalar potential given by (\ref{W})
in the limit of unbroken supersymmetry and, writing the adjoint fields in 
the fundamental basis\footnote{We define the fundamental representation
of $SU(N)$ $t_a$ such that $\Tr \(t_a t_b \) = {1\over 2} \delta_{ab} $ and 
$[t_a,t_b] = f_{abc} T_c $, while for the adjoint representation, e.g. 
$T^a_{ij} = f_{aij} $, we have $\Tr \(T_a T_b \) = N \delta_{ab} $.}
$\phi_i = \phi_i^a t_a $, it is given by:
\be
V = {\lambda^2 \over 4} | \phi_1^a \phi_2^a|^2 +  {\lambda^2 \over 4} |S|^2 
(| \phi_1^a |^2 + |\phi_2^a |^2) + {|D_a|^2 \over 2}
\label{scalpot}
\ee
where $S,\phi_i$ indicate now the scalar components of
the chiral multiplets, summation over $a$ is implicit and 
\be
D_a = i {g\over 2} f_{abc} \( \phi_1^{b*} \phi_1^c
+ \phi_2^{b*} \phi_2^c \)
\label{Dterm}
\ee
with $g$ denoting the $SU(N)$ gauge coupling. 

We see clearly that a flat direction exists for 
\bea
S &=& 0 \\
 \phi_1^a \phi_2^a &=& 0 \\
 f_{abc} \ \phi_i^{b*} \phi_i^c &=& 0.
\eea

This is not the most general case and other flat directions are present,
parametrized by gauge invariant polynomials \cite{fd}. 

For our purposes we will anyway consider the case when the inflaton is 
one of the components of the charged fields, i.e. we will take 
$\phi = Re \[ \phi_1^a\] $ 
for example and all the other fields driven to zero. Given some dominant 
non-zero component $\phi_1^a $ the potential is such that $S$ and 
$\phi_2^a$ are driven to zero by the effective mass term  
$\lambda^2 |\phi_1^a|^2 $. The gauge group $SU(N)$ is broken
by the inflaton v.e.v. and some combinations of the other 
charged fields components, depending on the rank of the 
residual gauge group, are eaten up by the gauge bosons
through the Higgs mechanism. For what concerns the other fields,
either they are driven to zero by the D term or they do not 
interact with the inflaton\footnote{In the case of 
$SU(2)$ we will have a $U(1)$ residual symmetry and 
two of the gauge bosons would acquire mass together with two
physical scalars and the corresponding fermions 
giving 2 massive vector multiplets. The remaining scalar
$Im \[ \phi_1^a\] $ will also obtain a large mass from the D-term.
For larger groups, it depends on the pattern of the 
breaking, e.g. $SU(5)$ could be broken to the SM group (with 12
massive gauge bosons) or to $SU(4)\times U(1)$ (with 8 massive gauge 
bosons) depending on which direction the inflaton field points. }.

Then for the field $\phi$ the potential reduces to zero and the only
contribution is from the soft susy breaking terms \cite{ssb}
\be
V_{ssb} = V_0 + m^2_S |S|^2 + m^2_1 | \phi_1^a |^2 + m^2_2 |\phi_2^a |^2
+ ( {Y \over 2} \lambda S  \phi_1^a \phi_2^a + h.c. )
\label{Vssb}
\ee
where $m_S, m_{1/2} $ are respectively the singlet and the charged fields
susy breaking masses and $Y$ denotes the supersymmetry breaking trilinear term.
From supergravity, we expect all the susy breaking scalar masses to be of 
order of $V_0$ and the trilinear parameter to be of order $V_0^{1/2}$. 
$V_0$ is a cosmological constant that is generated by some other sector 
of the theory and is cancelled in the true vacuum by the v.e.v. of 
some field in our sector, playing the role of the $\psi$. 
Notice that if we consider
the tree level potential, such field will have to develop a v.e.v. of
the order of the Planck scale, since $ V_0/|m^2_\psi| \simeq M^2_P$
reintroducing the Planck mass.

We see that along our flat direction, the potential reduces exactely 
to the form of eq. (\ref{vinf}). For writing the RGE improved potential, 
we will need to consider
the one loop renormalization group equations for our parameters.

\section{Renormalization Group Equations}

Following \cite{ma93}, we write down the equations for our particle content.
The gauge field strenght $\alpha = g^2/(4\pi) $ and the gaugino mass satisfy
\bea
{d\alpha\over dt} &=& {\beta\over 2\pi} \alpha^2 \label{RGEalpha}\\
{d\tilde m\over dt} &=&   {\beta\over 2\pi} \alpha \tilde m
\label{RGEmtilde} \eea
where $t= \ln (Q) $ is the renormalization scale and 
$\beta = - N$ in our case of $SU(N)$ with two matter superfields in the
adjoint representation ($\beta = -3N + n_{adj} N$).

This two equations are independent from the others and their 
solution is 
\bea
\alpha (t) &=& {\alpha_0 \over 1 - {\beta\over 2\pi} \alpha_0 t}
= {\alpha_0 \over 1 + \tilde \alpha_0 t}\\
\tilde m (t) &=& {\tilde m_0 \over \alpha_0} \alpha (t)
\eea
where $\tilde\alpha_0 = N \alpha_0/(2\pi) $.

For the Yukawa coupling, which we can always take real absorbing
its phase in the definition of the singlet field $S$, 
we have instead
\be
{d\lambda\over dt} = - N {\alpha\over \pi} \lambda + 
{\lambda\over 16\pi^2} (N^2+1) |\lambda |^2 
\label{RGEyukawa}
\ee
while the soft susy breaking masses follow the equations
\bea
{dm^2_S\over dt} &=& {N^2 -1 \over 16\pi^2} |\lambda|^2 \[ 2 m^2_S +
2 ( m^2_1 + m^2_2) + |Y|^2 \] 
\label{RGE-smass}\\
{dm^2_i\over dt} &=& {|\lambda |^2 \over 16\pi^2} \[ 2 m^2_i +
2 ( m^2_S + m^2_j) + |Y|^2 \] - {2 N \alpha\over \pi} \tilde m^2 
\label{RGE-inflmass}
\eea
where $i\neq j$, $m_S,m_i$ are respectively the masses of $S, \phi_i$
and $Y$ is the trilinear susy breaking term. 

We can simplify this system of first order equations by considering
instead than the charged particle masses, other variables, i.e.
\bea
m^2_{1-2} &=& m^2_1 - m^2_2 \\
m^2_{1-S} &=& m^2_1 - {1\over N^2-1} m^2_S ;
\eea
we can then recast the system into the form:
\bea
{dm^2_{1-2} \over dt} &=& 0 \\
{dm^2_{1-S} \over dt} &=& - {2 N \alpha\over \pi} \tilde m^2 
\label{RGE1Smass}\\
{dm^2_S\over dt} &=& {N^2 +1 \over 8\pi^2} |\lambda|^2   m^2_S + 
 {N^2 -1 \over 8\pi^2} |\lambda|^2 \[ 2 m^2_{1-S} - m^2_{1-2} + 
{|Y|^2\over 2} \].
\label{RGESmass} 
\eea

The trilinear term will have instead the equation
\be
{dY\over dt} = {1\over 32\pi^2} (N^2+1) Y |\lambda|^2 + 
{2\over\pi} N\alpha\tilde m .
\label{RGEtrilinear} 
\ee

These are a system of coupled differential equations.
We will in the next sections consider approximate solutions 
in three different cases and obtain the running inflaton mass.

\section{Small Yukawa coupling: $\lambda^2 \ll \alpha$}

This case is the simpler and has been already considered in a model
independent form in \cite{clr98}. In this approximation we can neglect the
$\lambda^2$ terms compared with the $\alpha$ ones and inflaton mass
running does not depend on the Yukawa coupling.
We have then
\be
m^2_i (t)  = m^2_{i,0} - A_0 \[ 1- {1\over ( 1 + \tilde \alpha_0 t)^2}\] 
\label{mrun1}
\ee
where $A_0 =  2 \tilde m^2_0$. Notice that eq.(\ref{mrun1}) gives in 
general a solution of eq.(\ref{RGE1Smass}), in the case of non negligible 
Yukawa.

For this expression for the running mass, the potential has a form 
studied in \cite{clr98}: the inflaton mass becomes positive as $\phi$ 
decreases and near the region where it vanishes the flatness condition
\ref{eta} is satisfied.
We will not consider here the inflaton dynamics in this potential, 
but analyze
instead the condition for the end of inflation, namely the fact that
the critical value does depend on the Yukawa.

Assuming universality in our sector, i.e. that all the scalar 
masses $m^2_S, m^2_i$ are equal at the
Planck scale, we see that naturally the role of the $\psi$ field is
played by the singlet field $S$, whose mass will stay practically constant
and negative.

Two issues are to be considered here; first the fact that in the limit
$\lambda \rightarrow 0$ the critical value becomes very large and 
practically no inflation takes place along our flat direction.
We have that at tree level, from eq. (\ref{scalpot}),
\be
\phi^2_c = {- 4 m^2_{S,0} \over \lambda_0^2} 
\simeq {4 V_0 \over \lambda_0^2}
\label{phicrit}
\ee
and therefore a lower bound on the Yukawa couplings appears if
we ask to have sufficient inflation.

Following the discussion in \cite{clr98} we will consider that slow roll
inflation ends before the critical value is reached and consider the
bound on the Yukawa coupling coming from such condition,
neglecting for the moment the running of both $\lambda, m^2_S$
(we will check later the consistency of our assumption).
The end of slow roll is given by $\eta (\phi_{end}) =1$, i.e.
using (\ref{mrun1}) and (\ref{eta}),
\be
\phi^2_{end} \simeq \exp \[-{2\over \tilde \alpha_0} 
\(1-{1\over \sqrt{1+{V_0+|m^2_{1,0}| \over A_0}}}\) \] 
\label{phiend}
\ee
in Planck units; we have then the bound
\be
\lambda_0^2 \geq 4 V_0 \exp \[ {2\over \tilde \alpha_0} 
\( 1-{1\over \sqrt{1+{V_0+|m^2_{1,0}| \over A_0}} } \) \].
\label{boundlambda}
\ee

As we can see this bound is very sensitive to the value of the gauge
coupling and also $V_0$. For value of the parameters in the
acceptable range given by \cite{clr98}, we have for example,
for $|m^2_{1,0}| = V_0, A_0 = 2 V_0, V_0 = 10^{-32} $, 
\be
8\times 10^{-7} \leq  \lambda_0^2 \ll \alpha_0 = 0.01 {2\pi\over N} 
\ee
so that in this case an acceptable range of Yukawa couplings exists
where our approximation is valid and slow roll ends before the 
critical value.
We see anyway that the gauge coupling has to be not too small, otherwise 
the bound (\ref{boundlambda}) is not consistent with
our initial assumption $\lambda_0^2 \ll \alpha_0 $. We can find a
lower bound on $\alpha_0$ for any choice of the other parameters
from the inequality:
\be
{1\over 2} \tilde\alpha_0 \ln\( {\pi\tilde\alpha_0 \over 2 N V_0} \)
\geq 1-{1\over\sqrt{1+{V_0+|m^2_{1,0}| \over A_0}}}.
\ee

The second issue is to check how our initial assumption is 
modified by the running of the relevant quantities, which even if 
small is there.
To estimate it let us consider the approximate equation
\be
{d\lambda\over dt} = - N {\alpha\over \pi} \lambda;
\label{RGElambdasmall}
\ee
the solution to such equation is
\be
\lambda = \lambda_0 {\alpha^2\over \alpha_0^2}.
\label{lambdasmall}
\ee
We see therefore that
\be
{\lambda \over \alpha } = {\lambda_0\over \alpha_0} {\alpha \over \alpha_0}
\ee
so that out initial assumption $\lambda_0^2 \ll \alpha_0$ 
remains valid as long as we do not 
enter the region $\alpha \gg \alpha_0 $, which anyway will coincide to
the non perturbative region for reasonable values of $\alpha_0$.

Finally let us consider how the definition of the critical value
is modified by the running; we have already obtained the running of
the Yukawa coupling in eq. (\ref{lambdasmall}). We can see clearly
from eq. (\ref{RGESmass}) that the corrections to $m^2_S$ due to
the running are 
of order $\lambda_0^2$ or higher. To lowest order in $\lambda_0^2$
we have therefore
\be
\phi_c^2 = {- 4 m^2_{S,0} \over \lambda^2 (t_c)} 
\simeq { 4 V_0 \over \lambda_0^2 } (1 + \tilde \alpha_0 t_c)^4;
\ee
we see that the correction to the tree level relation is of
order $\alpha_0$ and therefore small in the
perturbative regime.

\section{Fixed point region: $ \lambda^2 \geq \alpha $}

Let us consider now the case when $\lambda^2 \simeq \alpha $
and we can assume that the Yukawa coupling is quickly driven
towards the fixed point solution
\be
\lambda^2 = {12\pi N\over N^2+1} \alpha .
\label{fixedpoint}
\ee

In this case the equation for the singlet mass and the trilinear term
become
\bea
{dm^2_S\over dt} &=& {3 N\over 2 \pi} \alpha m^2_S + 
{3 N ( N^2 -1) \over 2 \pi (N^2+1)} \alpha 
\[ 2 m^2_{1-S} -  m^2_{1-2} + {1\over 2} |Y|^2 \] \\
{dY\over dt} &=& {3 N\over 8\pi} \alpha Y  + 
{2 N \over\pi} \alpha\tilde m ;
\eea
they are easily solved, and we have for a real $Y_0$:
\bea
Y (t) &=& (Y_0 + {16\over 7} \tilde m_0 ) (1+\tilde\alpha_0 t)^{3/4} - 
{16\over 7} \tilde m_0 {1\over 1+\tilde\alpha_0 t} \\
m^2_S (t) &=&  \left[ m^2_{S,0} + (N^2-1) (B_0 + C_0 + D_0 + E_0) \right]
 \( 1+\tilde\alpha_0 t\)^3 - (N^2-1) B_0 \( 1+\tilde\alpha_0 t\)^{3/2} 
\nonumber \\
& & - (N^2-1) C_0 - {D_0 (N^2-1) \over (1+\tilde\alpha_0 t)^{1/4}} 
- { E_0 (N^2-1)\over \(1+\tilde\alpha_0 t\)^2} 
\eea
where
\bea
B_0 &=& {1\over N^2+1}\[ {16\over 7} \tilde m_0 + Y_0 \]^2 \\
C_0 &=& {1\over N^2+1} \[m^2_{1,0} + m^2_{2,0} -{2\over N^2-1} 
m^2_{S,0} - 4 \tilde m^2_0 \] \\ 
D_0 &=& - {192\over 91 (N^2+1)} \tilde m_0 
\[ {16\over 7} \tilde m_0 + Y_0 \] \\
E_0 &=& {972 \over 245 (N^2+1)} \tilde m^2_0
\eea
are constant given in terms of the initial values.

From this solutions it is easy to obtain the inflaton mass in our case
\bea
m^2_1 (t) &=& m^2_{1-S} (t) + {m^2_S (t)\over N^2-1}\\
&=& m^2_{1,0} -{m^2_{S,0}\over N^2-1} - A_0 - C_0 
+ {A_0-E_0\over ( 1 + \tilde \alpha_0 t)^2}  
-{D_0\over (1+\tilde \alpha_0 t)^{1/4}} \nonumber\\
&-& B_0 \( 1+\tilde\alpha_0 t\)^{3/2}
+ \( {m^2_{S,0} \over N^2-1} + B_0 + C_0 + D_0 + E_0\) 
\( 1+\tilde\alpha_0 t\)^3.
\label{mrun2}
\eea

\begin{figure}
\centering
\leavevmode\epsfysize=6.5cm \epsfbox{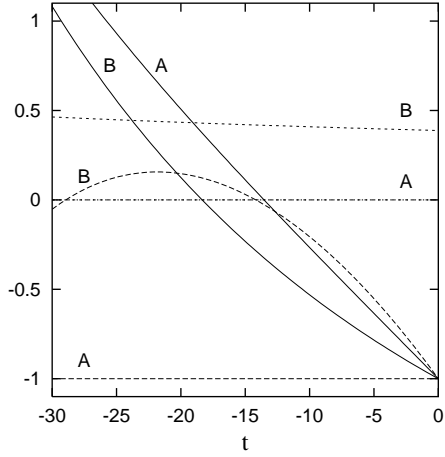}\\
\caption{\label{f:1} This figure shows the dependence of the masses
and lambda coupling on the scale $t=\ln(\phi)$ for universal masses
$- m^2_0$ at the Planck scale for $SU(2)$. We have taken in both cases 
$\tilde \alpha_0 = 0.01$, $\tilde m_0 = - Y_0 = m_0$ and rescaled the 
masses with respect to the initial absolute value.
The lines labelled $A$ refer to the small Yukawa coupling case with
$\lambda_0 = 10^{-5}$, while those labelled $B$ to the fixed point 
case $\lambda_0 = 0.388$. The full lines are the mass of the charged 
field $m^2_1(t)/m^2_0$, the dashed lines the singlet mass 
$m^2_S(t)/m^2_0$ and the dotted lines the Yukawa coupling $\lambda$.}
\end{figure}

We show the behaviour of the inflaton and singlet mass on Figure 1 for
the case of $SU(2)$ and universal masses at Planck scale.
Both the curves for the case of negligible Yukawa, eq. (\ref{mrun1}),
and fixed point solution, eq. (\ref{mrun2}), are shown for comparison.
Notice that the running is slightly 
weaker in the second case due to the presence
of the Yukawa and trilinear terms, but that the
inflaton mass tends to positive values also in this case.
Moreover the singlet mass $m^2_S $, after becoming positive 
in a small region, returns negative,
so that an hybrid end of inflation in this direction is always 
possible. 
The critical value in this case will be defined implicitly by
\be
\phi^2_c = {-4 m^2_S (t_c) \over \lambda (t_c)}.
\label{phicrit-run}
\ee

If we relax universality of the susy breaking masses at the Planck 
scale, taking $m^2_{2,0} \ll m^2_{1,0} $ and $m^2_{S,0}$ sufficiently 
positive, the role of the $\psi$ field can be played instead by the 
other charged field $\phi_2^a$ and the true vacuum would correspond 
to a broken gauge group.

\section{Large Yukawa coupling: $\lambda^2 \gg \alpha $}

In this case we can neglect the terms proportional to $\alpha$
with respect to those proportional to $\lambda^2$ and the
equations become similar to those for uncharged fields.
We can therefore consider at the same time the model where
$\phi_i$ are just two singlet fields substituting in the
following $ N^2 \rightarrow 2$. This substitution
amounts to consider only one degree of freedom instead of the
$N^2-1$ of a field in the adjoint representation of $SU(N)$.

The Yukawa coupling follows the equation:
\be
\lambda^2 = {\lambda^2_0 \over 1-{N^2+1 \over 2\pi^2} 
\lambda_0^2 t};
\ee
while the trilinear term is given by
\be
Y^2 =  {Y^2_0 \over \sqrt{1-{N^2+1 \over 2\pi^2} 
\lambda_0^2 t}}.
\ee
All the scalar masses have a behaviour similar to that
of $m^2_S$:
\bea
m^2_S (t) &=& {N^2-1\over N^2+1} \[(m^2_{S,0} + m^2_{1,0} +
m^2_{2,0} + Y^2_0 ) {1\over 1-{N^2+1 \over 2\pi^2} 
\lambda_0^2 t } \right. \nonumber  \\
& & \left. -Y_0^2 {1\over \sqrt{1-{N^2+1 \over 2\pi^2} 
\lambda_0^2 t}} - m^2_{1,0} - m^2_{2,0} + {2\over N^2-1}
m^2_{S,0} \] \\
m^2_i (t) &=& m^2_{i,0} + {1\over N^2-1} (m^2_S (t)-m^2_{S,0}).
\eea

Note that in case of universality and $SU(N)$, only the 
singlet mass changes sign due to the running; the other
charged particle masses run slower since they interact
with less particles. So a natural scenario is that 
where the role of the inflaton is played by the field $S$, 
while some other of the fields, whose mass stays negative, 
is $\psi$. Figure 2 displays the behaviour of the masses in
case of a common initial value.

\begin{figure}
\centering
\leavevmode\epsfysize=6.5cm \epsfbox{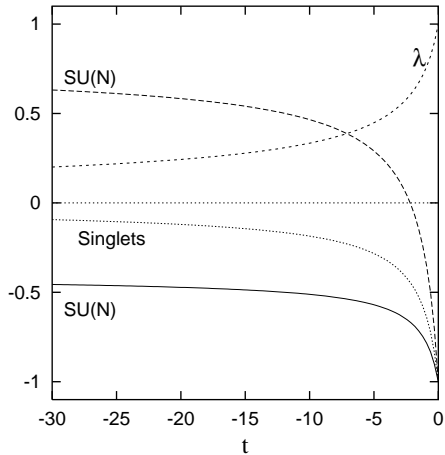}\\
\caption{\label{f:2} This figure shows the dependence of the Yukawa
coupling and masses on $t$ in the case when the gauge coupling is 
negligible and a universal mass scale is assumed at $M_{Pl}$.
We have taken $\lambda_0 = 1, Y_0 = - m_0$ and rescaled all the 
masses with
respect to the initial absolute value $m^2_0$. 
The curves labelled $SU(N)$ are related to the masses
of the singlet field (dashed line) and charged fields (full line)
in the $SU(N)$ model, while the line marked ``Singlets'' gives
the mass running for the 3 singlets model.}
\end{figure}

In the case of three singlet fields, all masses run similarly
and never change sign, therefore the only way of having an 
hybrid inflationary scenario, without invoking other fields and 
Yukawa couplings, would be to relax universality.
One possibility for inflation in this case, is to have negative 
initial masses and let one of them become positive like for
the $SU(N)$ case. Another option
is that initial different positive masses are driven negative, 
or very small for what regards the inflaton, 
by the Yukawa coupling like in the case of the 
radiative EW breaking in the MSSM. In such a picture not only would the
quantum corrections be responsible for the flattening of the
potential, but also for the triggering of the hybrid--type 
end of inflation. 

The last two possibilities are shown in figure 3 for specific
choices of the parameters.

\begin{figure}
\centering
\leavevmode\epsfysize=6.5cm \epsfbox{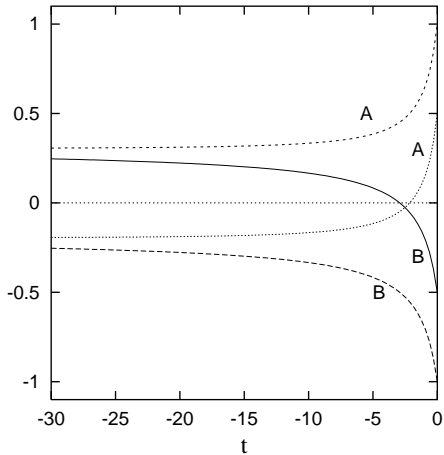}\\
\caption{\label{f:3} The running of the different
masses in case of non-universality of the initial mass scale for
$\lambda_0 =1, Y_0 = - m_0$ in the three singlet scenario. 
Case A: the initial masses are positive such that 
$m^2_{i,0} = m^2_{S,0}/2 $; the dashed line line gives the behaviour 
of the $S$ mass, while the dotted line of the other two masses.
Case B: the initial masses are negative and 
$m^2_{S,0} = m^2_{i,0}/2 $; the full line is $m^2_S (t)$ while the
dashed line is $m^2_i (t)$.}
\end{figure}

\section{Observational constraints}

The spectral index behaviour in the running mass models is more
easily understood using the linear approximation \cite{cl98}.
Cosmological scales must leave the horizon when the inflaton
v.e.v. lies in the flat region of the potential, i.e. in the
neighborhood of the point where $\epsilon$ vanish, 
$\epsilon (t_\star) = 0$ and $\ln (\phi_\star) = t_\star$.
As described in \cite{cl98}, all the observational constraints
on the spectral index and COBE normalization can be cast in a
simple form considering the linear approximation for the
running inflaton mass around $\phi_\star$ 
and defining the three parameters:
\bea
c &=& - {d m^2_\phi \over dt}|_{t=t_\star} = 2 m^2_\phi (t_\star) \\
\tau &=& - |c| \ln(\phi_\star) \\
\sigma &=& \lim_{\phi\rightarrow\phi_\star} c e^{c N(\phi)} 
\ln (\phi_\star/\phi),
\eea
where $N(\phi) $ is the number of e-folds from $\phi$ to $\phi_{end}$
in the slow roll approximation:
\be
N(\phi) = \int^\phi_{\phi_{end}} {V \over V'} d\phi.
\ee

Computing this quantities in the specific cases would make 
possible to give the behaviour of the spectral index
scale dependence and the bound on $V_0$ from COBE normalization
and restrict the initial parameter space. 
We will not do this here for our particular model, but refer to
\cite{cl98} for a model independent analysis on the allowed
parameter space for $c,\tau$ and $\sigma$. Let us note that in
the language of that reference, the case of a negative inflaton mass
at high scales corresponds to models (i) and (ii), while a positive
inflaton mass at $M_P$, as mentioned in the last section, leads to
models (iii) and (iv).
 
A clear experimental signature of this kind of running mass
models is anyway the scale dependence of the spectral index
$n$, 
\be
{n-1\over 2} = \sigma e^{-c N} - c
\ee
that will be within the reach of the future Planck 
satellite measurements \cite{planck}
in a large part of the parameter space.

\section{Conclusions}

We have described a particular model of hybrid inflation with
a running mass and found that an interesting regime for
inflation is present for any hierarchy of the initial couplings.

In most cases the potential is flattened by the gauge 
interaction and a charged inflaton is needed, but for large
Yukawa coupling also this kind of corrections can be
sufficient to give a small $\eta $ and a singlet inflaton
can be considered.

Further investigations will be required for delimiting precisely
the region in 
parameter space compatible with the present observational constraints
and determine the naturalness of this scenarios. One point we would like 
to make is that a sizable region has been found for the case of a 
negligible Yukawa coupling \cite{clr98}. The introduction of
a Yukawa coupling near the fixed point solution, as discussed in 
section 6, should enlarge such region since $\lambda $ slows the 
running of the inflaton mass and therefore weakens the scale 
dependence of the spectral index.
For what concerns the large Yukawa and the singlet inflaton cases,
a more detailed analysis is needed, in particular in order to see
how large $\lambda$ has to be to flatten sufficiently 
the potential. Notice that contrary to what happens in the usual 
inflationary picture, in our case the inflaton has to have sizeable
 couplings to other fields in order for the one loop correction to
 flatten the potential. 

Another issue we would like to mention here is the identification
of the sector the inflaton and the other fields belong to; since no
matter fields in the adjoint representation appear in the
MSSM, we are obliged to extend its particle content.

One possibility could be to identify our gauge group with colour
$SU(3)$ and assume that the colour symmetry is broken during inflation
to be restored at the end. Since at that point our adjoint fields would
acquire a mass of order $M_P$ from the $S$ v.e.v., such scenario would 
not modify our low energy phenomenology. In that case all the
particle of the MSSM would have to be taken into account and the 
RGE equations would be different from those studied here. In particular
the colour gauge group would be non--asymptotically free due to the
quarks contribution to $\beta$.
A successful inflationary scenario can be constructed anyway in the 
non--asymptotically free case \cite{cl98}, but the picture would be 
certainly more complicated than described here due to the many fields 
involved and less attractive from the theoretical point of view.

Otherwise, we could think of accommodate our fields into some
Gran Unified Theory, like $SU(5)$, or in some hidden sector.
In the first case, the ``right'' symmetry breaking has to be
considered in order to end in the MSSM minimum through
a charged $\psi$ field and care has
to be taken to insure that all the GUT particles masses
are at the right scale. It would be probably difficult to
realize such a picture without some sort of tuning.

In the second case instead, the model is certainly less constraint. 
Reheating or preheating could in this case be
difficult to realize if the only interaction between the hidden and 
visible sector is gravitational. In this case anyway,
the role of the $\psi$ field
could be played by $S$ and reheating the universe to our
visible sector could proceed through interactions of this 
singlet field with SM fields. If such singlet could
be identified with a bulk modulus that could be achieved
and a v.e.v. of the order of the Planck mass would be
natural.

All this possibilities will be the subject of further studies.

Let us conclude here recalling that in any case the 
experimental signature of this kind of models, i.e. the scale
dependence of the spectral index, is usually non negligible
and within the reach of the next satellite experiments.

\section*{Acknowledgements}

The author would like to thank D. Lyth for many discussions on
this subject, encouragement and comments on the manuscript and
L. Roszkowski for discussions and encouragement; she is also grateful 
to M. Bastero-Gil, 
E. Copeland, A. C. Davis, S. Davis, X. Meng, G. Ross and S. Sarkar for 
discussions during the XIXth UK Institute for Theoretical High Energy 
Physicists in Oxford. The author is supported by PPARC grant GR/L40649.

\def\PLB#1#2#3{Phys. Lett. {\bf B#1}, #3 (19#2) }
\def\PLBold#1#2#3{Phys. Lett. {\bf#1B} (19#2) #3}
\def\PRD#1#2#3{Phys. Rev. {\bf D#1}, #3 (19#2) }
\def\PRL#1#2#3{Phys. Rev. Lett. {\bf#1} (19#2) #3}
\def\PRT#1#2#3{Phys. Rep. {\bf#1} (19#2) #3}
\def\ARAA#1#2#3{Ann. Rev. Astron. Astrophys. {\bf#1} (19#2) #3}
\def\ARNP#1#2#3{Ann. Rev. Nucl. Part. Sci. {\bf#1} (19#2) #3}
\def\mpl#1#2#3{Mod. Phys. Lett. {\bf #1} (19#2) #3}
\def\ZPC#1#2#3{Zeit. f\"ur Physik {\bf C#1} (19#2) #3}
\def\APJ#1#2#3{Ap. J. {\bf #1} (19#2) #3}
\def\AP#1#2#3{{Ann. Phys. } {\bf #1} (19#2) #3}
\def\RMP#1#2#3{{Rev. Mod. Phys. } {\bf #1} (19#2) #3}
\def\CMP#1#2#3{{Comm. Math. Phys. } {\bf #1} (19#2) #3}


\begin{thebibliography}{999}

\bibitem{ly98} D. H. Lyth and A. Riotto, hep-ph/9807278.

\bibitem{cllsw}  E. J. Copeland, A. R. Liddle, D. H. Lyth, 
E. D. Stewart and D. Wands, Phys. Rev. {\bf D49}, 6410 (1994).

\bibitem{ewanloop1} E. D. Stewart, Phys. Lett. {\bf B391}, 34 (1997).

\bibitem{ewanloop2}  E. D. Stewart,
Phys. Rev. {\bf D56}, 2019 (1997).

\bibitem{clr98} L. Covi, D. H. Lyth and L. Roszkowski,
hep-ph/9809310.      

\bibitem{cl98} L. Covi and D. H. Lyth, hep-ph/9809562.

\bibitem{ssb} D. Bailin and A. Love {\it Supersymmetric Gauge Field 
Theory and String Theory }, Istitute of Physics Publishing, Bristol 1994;

H. P. Nilles, Phys. Rep. 110, 1-162 (1984).

\bibitem{ma93} S. T. Martin and M. T. Vaughn, Phys. Rev. {\bf D50},
2282 (1994).

\bibitem{fd} M. A. Luty and W. Taylor, Phys. Rev. {\bf  D53}, 
3399 (1996).

\bibitem{planck} home page at {\tt http://astro.estec.esa.nl/Planck}

\end{thebibliography}
\end{document}